\documentclass[a4paper,11pt]{article}
\usepackage{pos}

\usepackage{sidecap}
\sidecaptionvpos{figure}{t}

\usepackage{ulem}

\title{Update on the gradient flow scale\\ on the 2+1+1 HISQ ensembles}
\ShortTitle{Gradient flow scale}

\author*[a,b]{Alexei~Bazavov}
\author[c]{Claude~Bernard}
\author[d]{Carleton~E.~DeTar}
\author[e,f]{Aida~X.~El-Khadra}
\author[g]{Elvira~G\'{a}miz}
\author[h]{Steven~Gottlieb}
\author[i]{Anthony~V.~Grebe}
\author[j]{Urs~M.~Heller}
\author[k]{William~I.~Jay}
\author[i]{Andreas~S.~Kronfeld}
\author[k,l]{Yin~Lin}

\affiliation[a]{Department of Computational Mathematics,
Science, Michigan State University, East Lansing, Michigan 48824, USA}
\affiliation[b]{Department of Physics and Astronomy,
Michigan State University, East Lansing, Michigan 48824, USA}

\affiliation[c]{Department of Physics, Washington University, St. Louis, Missouri 63130, USA}

\affiliation[d]{Department of Physics and Astronomy, University of Utah,
	Salt Lake City, Utah 84112, USA}

\affiliation[e]{Department of Physics, University of Illinois, Urbana, Illinois 61801, USA}
\affiliation[f]{Illinois Center for Advanced Studies of the Universe, University of Illinois, Urbana, Illinois 61801, USA}

\affiliation[g]{CAFPE and Departamento de Fisica Te\'{o}rica y del Cosmos,
	Universidad de Granada, E-18071 Granada, Spain}

\affiliation[h]{Department of Physics, Indiana University, Bloomington, Indiana 47405 USA}

\affiliation[i]{Theory Division, Fermi National Accelerator Laboratory, Batavia, Illinois 60510, USA}

\affiliation[j]{American Physical Society, Hauppauge, New York 11788, USA}

\affiliation[k]{Center for Theoretical Physics, Massachusetts Institute of Technology, Cambridge,
	Massachusets 02139, USA}

\affiliation[l]{The NSF AI Institute for Artificial Intelligence and Fundamental Interactions}

\abstract{We report on the ongoing effort of improving the determination of the gradient flow scale on the (2+1+1)-flavor HISQ ensembles generated by the MILC collaboration. We compute the scales $\sqrt{t_0}/a$ and $w_0/a$ with the Wilson and Symanzik flow using three discretizations for the action density: clover, Wilson and tree-level Symanzik improved. For the absolute scale setting, we intend to employ the $\Omega$-baryon mass, but are also using the pion decay constant while the $\Omega$-mass calculations are in progress.\\
	\\
	FERMILAB-CONF-23-0833-T\\
	MIT-CTP/5663
}

\FullConference{The 40th International Symposium on Lattice Field Theory (Lattice 2023)\\
July 31st - August 4th, 2023\\
Fermi National Accelerator Laboratory\\}

\begin{document}
\maketitle

\section{Introduction}

For lattice-QCD precision projects such as the calculation of the hadronic contribution to  the anomalous magnetic moment of the muon, a major contribution to the final uncertainty comes from the determination of the lattice spacing (see, for instance, Ref.~\cite{FermilabLatticeHPQCD:2023jof}). A convenient procedure for determining the relative lattice scale with high statistical precision is the gradient flow introduced in Ref.~\cite{Luscher:2010iy}. For the absolute scale, several quantities have been used in the past including the pion or kaon decay constants. However, the $\Omega$-baryon mass, which is measured experimentally to about 0.02\% precision, has recently become a more popular choice for precision lattice projects.

The MILC collaboration has been generating a library of (2+1+1)-flavor ensembles with the Highly Improved Staggered Quark (HISQ) action~\cite{Follana:2006rc} for over a decade now~\cite{MILC:2010pul,MILC:2012znn}. The available lattice spacings range from $0.15$ down to $0.03$ fm and the quark masses from approximately $300$~MeV down to the physical pion mass. The initial tuning was done with the $r_0/r_1$ scale and later improved with the $f_\pi$ scale. In 2015, the $w_0$ scale was calculated on the existing ensembles~\cite{MILC:2015tqx}. Since then, more ensembles have been added, and some physical mass ensembles were regenerated with more precisely tuned quark masses. Therefore, we are revisiting $w_0$-scale setting on the MILC HISQ ensembles. Improvements in this calculation include using a different integrator~\cite{LSCF2021,Bazavov:2021pik}, two types of gradient flow (Wilson and Symanzik), three different discretizations of the observable, and use of the $\Omega$-baryon mass for the absolute scale setting. Recent developments, \textit{e.g.}, Refs.~\cite{Lin:2019pia,Hughes:2019ico,Lin:2020wko}, allow us to use a fully unitary formalism where the valence quarks are also HISQ.
Recently, the CalLat collaboration computed the physical $w_0$ scale with the $\Omega$-baryon mass input using most of the MILC HISQ publicly available   ensembles including the ones with heavier-than-physical pion mass and a few additional HISQ ensembles generated by them~\cite{Miller:2020evg}. The main difference with our work is that they used the M\"obius domain-wall fermion action for the valence quarks and that we compute the $\Omega$-baryon mass only on the physical pion mass ensembles.

\section{Gradient flow}

\subsection{Definitions}

The gradient flow~\cite{Luscher:2010iy} is a smoothing procedure where a gauge field configuration is evolved in a fictitious time $t$ according to a diffusion equation,
\begin{equation}
\frac{dV_{x,\,\mu}}{dt}=-\left\{\partial_{x,\,\mu}S^f(t)\right\}V_{x,\,\mu},
\,\,\,\,\,\,V_{x,\,\mu}(t=0)=U_{x,\,\mu}\,.
\label{eq_flow}
\end{equation}
In this project, for the flow action $S^f$, we use both the Wilson and tree-level improved Symanzik actions. To set the lattice spacing, one evaluates a flowed observable, a popular choice being the action density $S^o$ itself, at flow time $t=t_0$ or $t=w_0^2$ where the corresponding condition is satisfied:
\begin{equation}
\left.t^2\langle S^o(t)\rangle\right|_{t=t_0}=0.3\,\,\,\,\,\,\,\,\,\,\,
\mbox{or}\,\,\,\,\,\,\,\,\,\,\,
\left[t\frac{d}{dt}t^2\langle S^o(t)\rangle\right]_{t=w_0^2}=0.3\,.
\end{equation}
The major practical difference between the resulting $\sqrt{t_0}$ or $w_0$ scales is higher suppression of the discretization effects in $w_0$, as was first noted in Ref.~\cite{BMW:2012hcm}.
Moreover, the discretization effects at the tree level that result from a specific combination of the gauge action $S^g$ (with HISQ we use a one-loop Symanzik-improved action), flow action $S^f$ and observable discretization $S^o$ were calculated in Ref.~\cite{Fodor:2014cpa} and can be canceled by replacing the observable with
\begin{equation}
	t^2 S^o(t)\,\,\,\rightarrow\,\,\,
	\frac{\displaystyle t^2 S^o(t)}{
		\displaystyle
		1+\sum_{m=1}^4 C_m(a^{2m}/t^m)}\,.
	\label{eq_corr}
\end{equation}
For the observable, we use Wilson (W), tree-level Symanzik-improved (S) and clover (C) discretization of the action density. In what follows, three letter abbreviations refer to the combination of the gauge, flow, and observable actions, \textit{e.g.}, the SSC (Symanzik-Symanzik-Clover) combination was the only one used in our previous work~\cite{MILC:2015tqx}.

\subsection{Lattice setup}

In general, computation of the gradient flow for a given ensemble requires a smaller computational effort than other typical calculations, \textit{e.g.}, hadron spectrum measurements. However, given the diffusive nature of the flow equation~(\ref{eq_flow}) the flow time is actually measured in units of lattice spacing squared. This means that for finer ensembles the flow needs to be run significantly longer (in lattice units) to reach the $t_0$ or $w_0^2$ scales. For this reason, we explored a number of integration schemes that would allow the flow integration to have the largest possible step size. We use a fourth-order commutator-free low-storage Lie group integrator described in Refs.~\cite{LSCF2021,Bazavov:2021pik}. To completely control the systematic errors from the flow integration, we perform all measurements with two step sizes $\Delta t=1/20$ and $\Delta t=1/40$.
We observe that the integration is stable for all ensembles and the systematic error is always a few orders of magnitude smaller than the statistical error.

\begin{figure}[h]
	\begin{center}
		\includegraphics[width=0.49\textwidth]{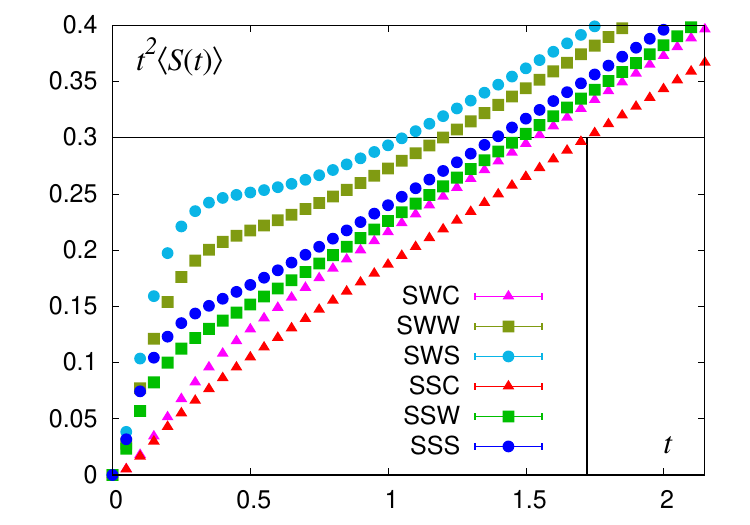}\hfill
		\includegraphics[width=0.49\textwidth]{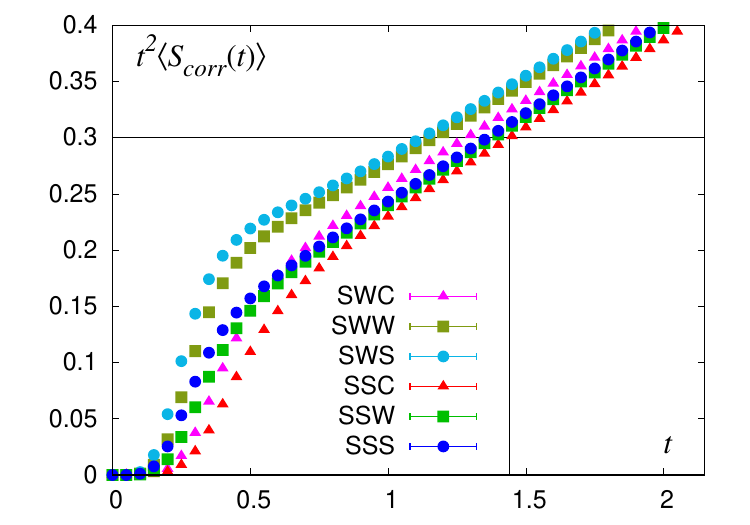}
		\includegraphics[width=0.49\textwidth]{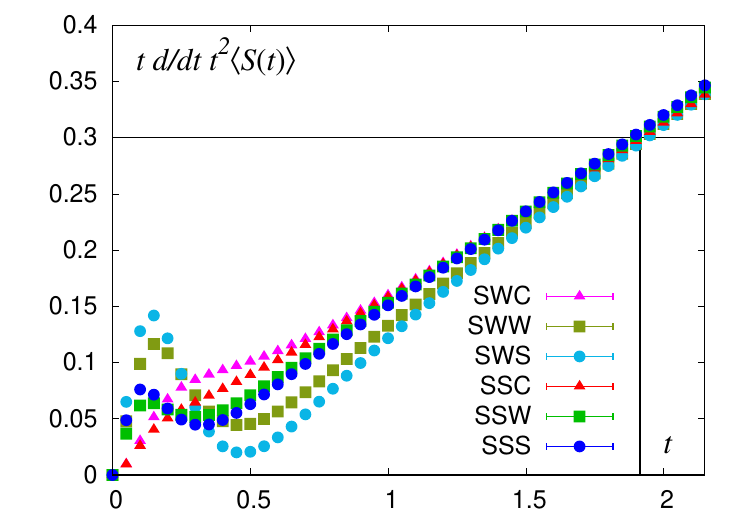}\hfill
		\includegraphics[width=0.49\textwidth]{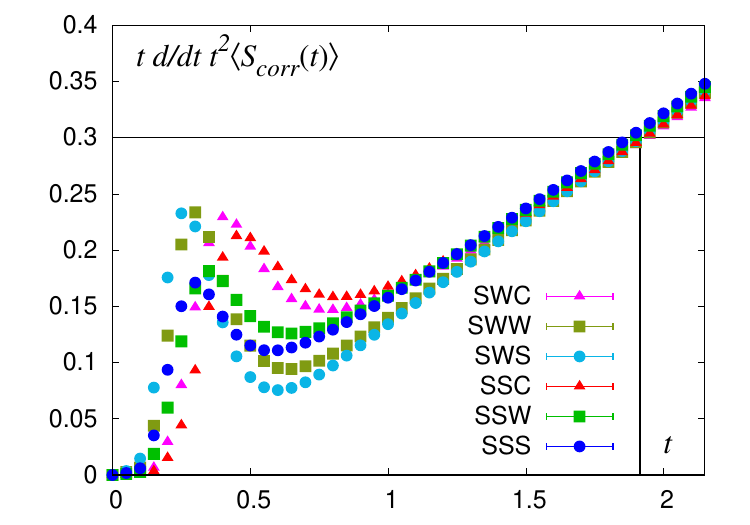}
		\caption{The flow time dependence (in lattice units) of different combinations of the flow and observable. Top panels show the observables used to extract the $\sqrt{t_0}$ scale and the bottom panels the $w_0$ scale. Left panels show the original observables, while right panels show the effect of applying tree-level corrections for discretization effects, Eq.~(\ref{eq_corr}). The statistical errors are smaller than the symbol sizes.\label{fig_flows_a12}}
	\end{center}
\end{figure}

We illustrate how different choices influence the discretization effects in Fig.~\ref{fig_flows_a12} for the 300~MeV pion, $a=0.12$~fm ensemble. In all panels, we show six different combinations of the flow/observable (with the gauge action being fixed at ``S''). In the top left panel of Fig.~\ref{fig_flows_a12}, the flows for the $\sqrt{t_0}$ scale are shown and the spread of different curves crossing the 0.3 line hints at relatively large discretization effects. The vertical line shows the scale that corresponds to the SSC combination that was used in our 2015 work~\cite{MILC:2015tqx}. In the top right panel, we show the effect of applying the corrections, Eq.~(\ref{eq_corr}), to the observable. The decrease in the spread between different flow/observable combinations shows some suppression of the discretization effects, but they are still quite large for the $\sqrt{t_0}$ scale at $a=0.12$~fm. Moving to the bottom left panel of Fig.~\ref{fig_flows_a12}, we find that the $w_0$ scale has drastically reduced discretization effects, and applying the corrections to $w_0$, bottom right panel of Fig.~\ref{fig_flows_a12}, has a relatively small effect. Note also that applying the corrections to the derivative ($w_0$) scale tends to amplify the artifacts at early flow time. The effect of applying the corrections to the $\sqrt{t_0}$ and $w_0$ scale is similar for the finer ensembles, so we do not show them here.

\subsection{Autocorrelations}

The scale $\sqrt{t_0}$ or $w_0$ became a popular choice due to a relatively small computational cost and high statistical precision that can be achieved with either quantity. However, the smearing introduced by the flow, in general, induces longer autocorrelation times for these quantities, compared to $r_0$ or even $f_\pi$ scales.

The autocovariance function for an observable $O$ is defined as
\begin{equation}
C(t_{MC})\equiv \langle({\cal O}_i-\langle{\cal O}_i\rangle)({\cal O}_j-\langle{\cal O}_j\rangle)\rangle
=\langle{\cal O}_0{\cal O}_{t_{MC}}\rangle-\langle{\cal O}\rangle^2,\,\,\,\,\,t_{MC}=|i-j|,
\end{equation}
where $i$, $j$ index the Monte Carlo time series. The autocorrelation function is simply $c(t_{MC})\equiv C(t_{MC})/C(0)$. The estimator of the integrated autocorrelation time with the window method is
\begin{equation}
\bar\tau_{int}(t_{MC})=1+2\sum_{t'=1}^{t_{MC}} c(t').
\label{eq_tauint}
\end{equation}
In general, the quantity (\ref{eq_tauint}) is expected to reach a plateau, but in practice this is complicated by the fact that the variance of $\bar\tau_{int}(t_{MC})$ diverges. To reliably estimate the integrated autocorrelation time requires significant statistics.
To propagate the statistical errors on the autocorrelation function and the integrated autocorrelation time, we use a block jackknife resampling with 20 and 40 blocks to check for the dependence on the block size.

If one assumes that the autocorrelation function is dominated by a single exponential, \textit{i.e.}, \begin{equation}
c(t_{MC})=\exp\{-t_{MC}/\tau^1_{MC}\}\,,
\label{eq_fit_ac}
\end{equation}
the integrated autocorrelation time is given by the sum of the geometric series
\begin{equation}
\tau^1_{int}=\frac{\exp\{1/\tau^1_{MC}\}+1}{\exp\{1/\tau^1_{MC}\}-1}.
\label{eq_tau1}
\end{equation}
This provides an alternative way of determining the integrated autocorrelation time by fitting the early time behavior of the autocorrelation function that can be determined with higher statistical precision.

\begin{figure}[h]
	\begin{center}
		\includegraphics[width=0.49\textwidth]{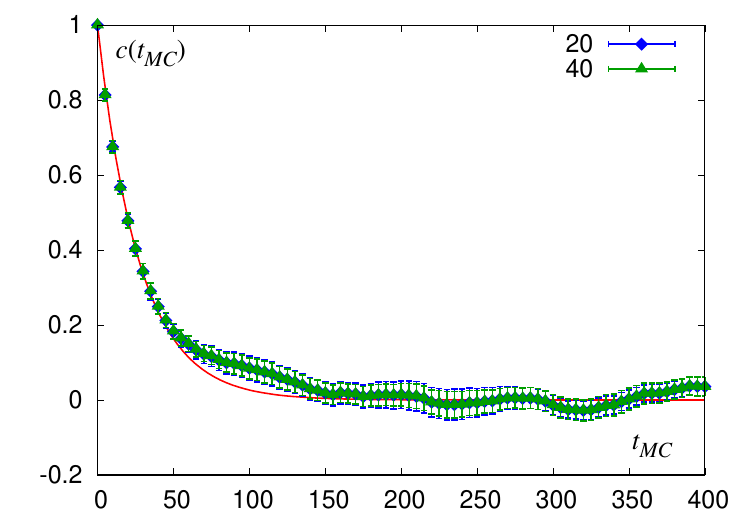}\hfill
		\includegraphics[width=0.49\textwidth]{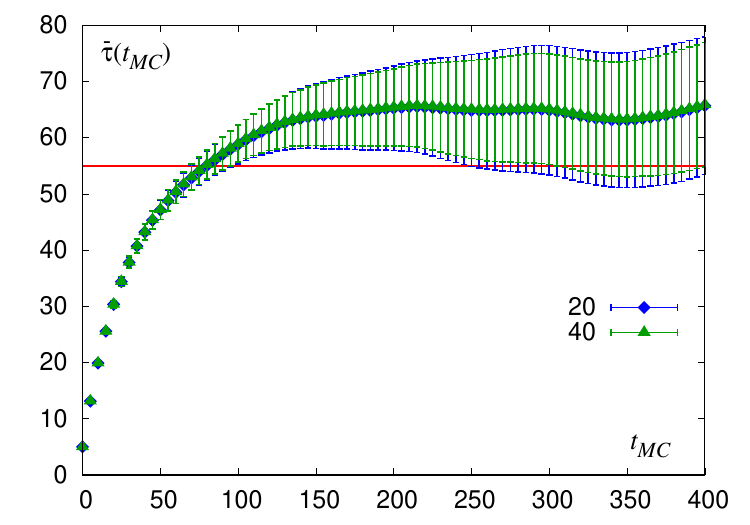}
		\caption{The autocorrelation function (left) and the integrated autocorrelation time (right) measured with the window method for $a=0.12$~fm ensemble vs.\ the simulation time in molecular dynamics time units. Different colors correspond to varying the block size in the jackknife procedure.
		\label{fig_tau_a12}}
	\end{center}
\end{figure}
For the retuned $a=0.12$~fm physical mass MILC HISQ ensemble, a large data set of 45,000 molecular dynamics time units has been accumulated (with lattices saved every 5 time units). 
To study autocorrelations we take the clover action density for the observable at time $t=w_0^2$,
$O\equiv S^o(t=w_0^2)$, \textit{i.e.}, the quantity from whose ensemble average the $w_0$ scale is determined.
In the left panel of Fig.~\ref{fig_tau_a12}, we show the normalized autocorrelation function $c(t_{MC})$ for that ensemble together with the fit to the form~(\ref{eq_fit_ac}) versus the Monte Carlo time in molecular dynamics time units. In the right panel of the figure, we show the integrated autocorrelation time $\bar\tau_{int}(t_{MC})$ computed with the window method together with the $\tau^1_{int}=55\pm3$ computed from assuming a single-exponential decay~(\ref{eq_tau1}). One can see that from such a large ensemble the autocorrelation function can be determined with enough statistical precision, and as a result one gets a reliable estimate of the integrated autocorrelation time. The value calculated from the single-exponential decay is reasonably consistent with the plateau.
Conservatively, we can estimate that the integrated autocorrelation time is below 80 molecular dynamics time units, and the ensemble has at least 45,000/80 $\sim$ 560 statistically independent samples as far as $w_0$ is concerned.

\begin{figure}[h]
	\begin{center}
		\includegraphics[width=0.49\textwidth]{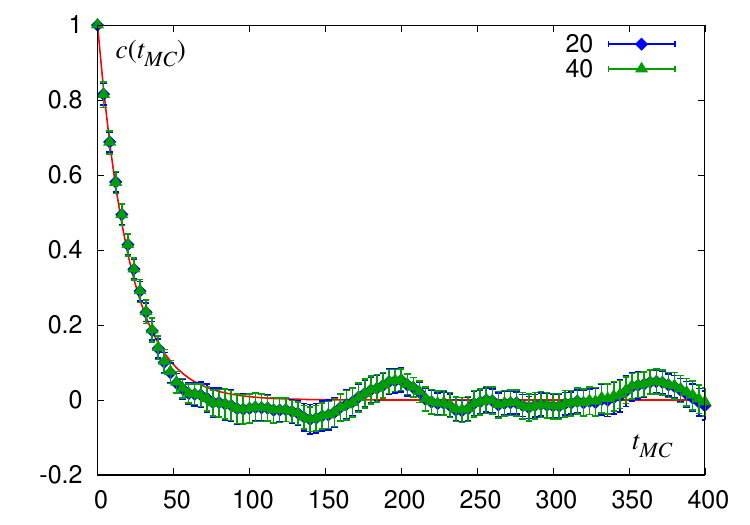}\hfill
		\includegraphics[width=0.49\textwidth]{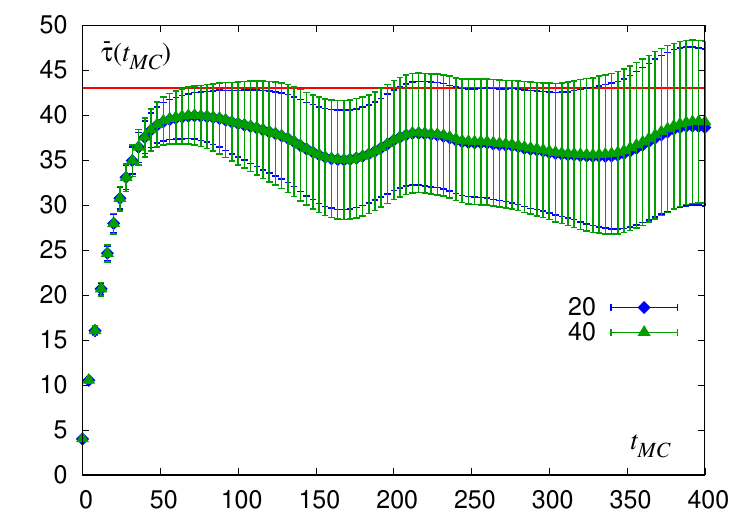}
		\caption{The autocorrelation function (left) and the integrated autocorrelation time (right) measured with the window method for $a=0.09$~fm ensemble vs.\ the simulation time in molecular dynamics time units. Different colors correspond to varying the block size in the jackknife procedure.
		\label{fig_tau_a09}}
	\end{center}
\end{figure}
Another high-statistics MILC HISQ ensemble at finer lattice spacing is the $a=0.09$~fm physical-mass ensemble with about 20,000 molecular dynamics time units. Similarly to the previous case, we look at the clover action density at the flow time $t=w_0^2$. Its autocorrelation function is shown in the left panel and the integrated autocorrelation time in the right panel of Fig.~\ref{fig_tau_a09}. The integrated autocorrelation time determined from the single-exponential decay is $\tau^1_{int}=43\pm3$ molecular dynamics time units and is consistent with the plateau achieved in $\bar\tau_{int}(t_{MC})$.

\begin{figure}[h]
	\begin{center}
		\includegraphics[width=0.49\textwidth]{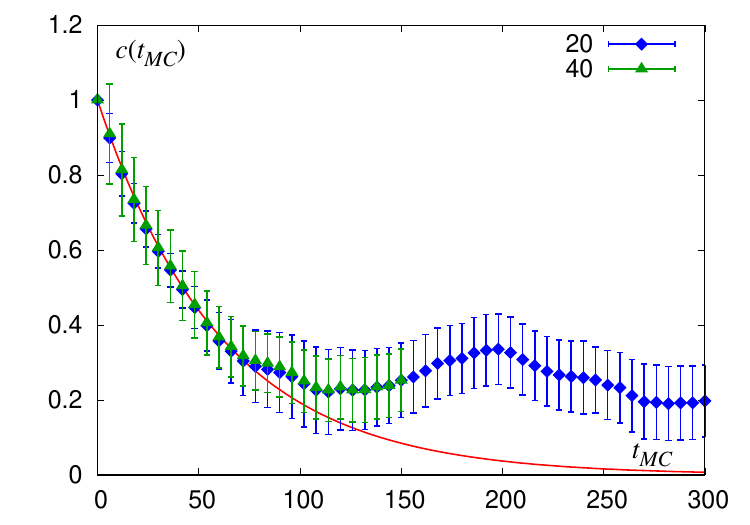}\hfill
		\includegraphics[width=0.49\textwidth]{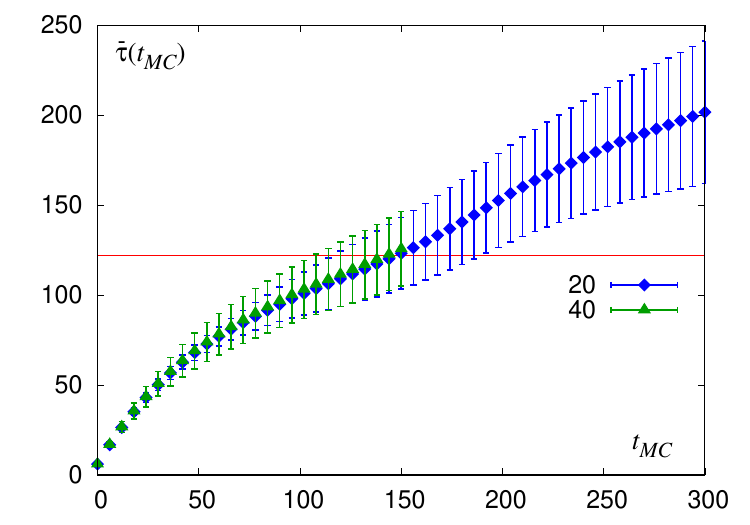}
		\caption{The autocorrelation function (left) and the integrated autocorrelation time (right) estimated with the window method for $a=0.06$~fm ensemble vs.\ the simulation time in molecular dynamics time units. Different colors correspond to varying the block size in the jackknife procedure. Note that with 40 blocks the autocorrelation function is computed only up to $t_{MC}=6,000/40=150$ molecular dynamics time units.
		\label{fig_tau_a06}}
	\end{center}
\end{figure}
We now turn to a finer, $a=0.06$~fm, 300~MeV pion mass MILC HISQ ensemble that contains 6,000 molecular dynamics time units (1,000 configurations separated by 6 time units). We expect that when decreasing the lattice spacing the integrated autocorrelation time for $w_0$ related observables will, in general, grow (in molecular dynamics time units). As before, we take the clover action density at $t=w_0^2$ as an observable and show its autocorrelation function in the left panel and the integrated autocorrelation time in the right panel of Fig.~\ref{fig_tau_a06}. Compared to the two previous cases, the errors on the autocorrelation function are much larger and the integrated autocorrelation time does not reach a plateau. It is plausible, based on our experience with the other ensembles, that the increase in the autocorrelation function around 200 time units is a statistical fluctuation. If the Monte Carlo statistics were significantly increased, the autocorrelation function (at least visually) would become more consistent with the single-exponential form, as was observed for the two large-statistics ensembles discussed previously. In that case, the integrated autocorrelation time might plateau around 120--150 time units, becoming consistent with the estimate $\tau^1_{int}=122\pm31$ time units. However, at present, the data do not allow one to rule out larger integrated autocorrelation times up to 250 time units. Conservatively, we can estimate that the $a=0.06$~fm 300~MeV pion MILC HISQ ensemble contains from 6,000/250 = 24 to 6,000/120 = 50 statistically independent samples.

Overall, we conclude that for the ensembles with the lattice spacing $a>0.06$~fm the integrated autocorrelation time for the gradient flow scale observables is on the order of 50 time units or below (it is mainly sensitive to the lattice spacing and less to the values of the quark masses) while for $\le 0.06$~fm ensembles the integrated autocorrelation time may be significant, and one has to exercise care when estimating the true variance of the $w_0$ scale.

We have currently completed the flow measurements on all existing MILC HISQ ensembles, and the final analysis is ongoing. The details will be presented in a forthcoming publication.

\section{Staggered baryons}

For the absolute scale setting, we have been computing the $\Omega$-baryon mass on the physical pion mass MILC HISQ ensembles with lattice spacings $a\approx 0.15$, 0.12, 0.09, and 0.06~fm. On each ensemble, we construct the full set of operators that interpolate to different staggered fermion tastes of $\Omega$ baryons as described in Ref.~\cite{Bailey:2006zn}. 
We use three interpolators here.
Two of three operators are mixed and interpolate to the same two taste-split $\Omega$ baryons in the ground state; on the other hand, the remaining operator interpolates to a single $\Omega$ baryon. 
In the continuum limit, the taste-splitting mass differences disappear and a single $\Omega$-baryon mass is obtained.
\begin{figure}[h]
	\begin{center}
		\includegraphics[width=0.49\textwidth]{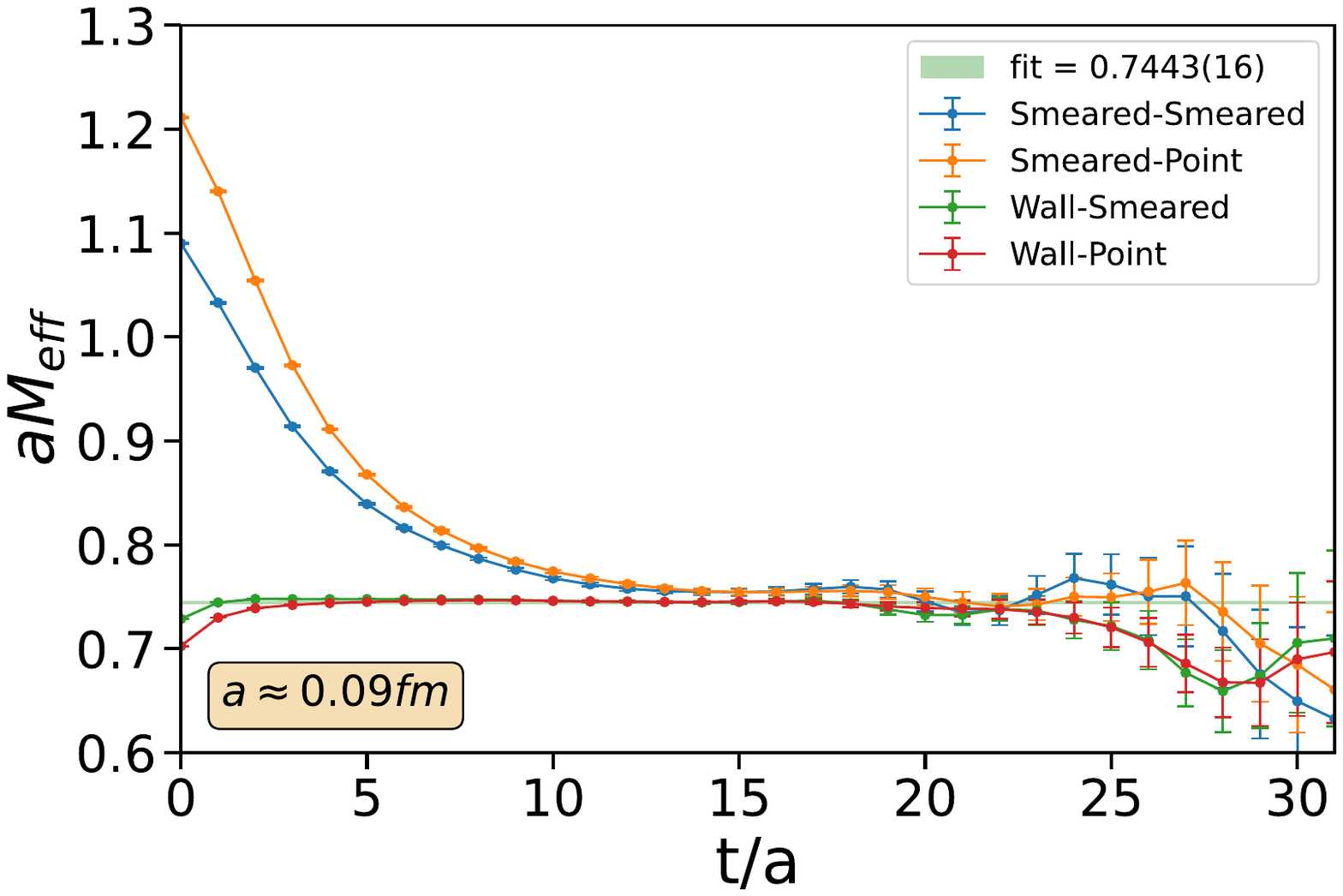}\hfill
		\includegraphics[width=0.49\textwidth]{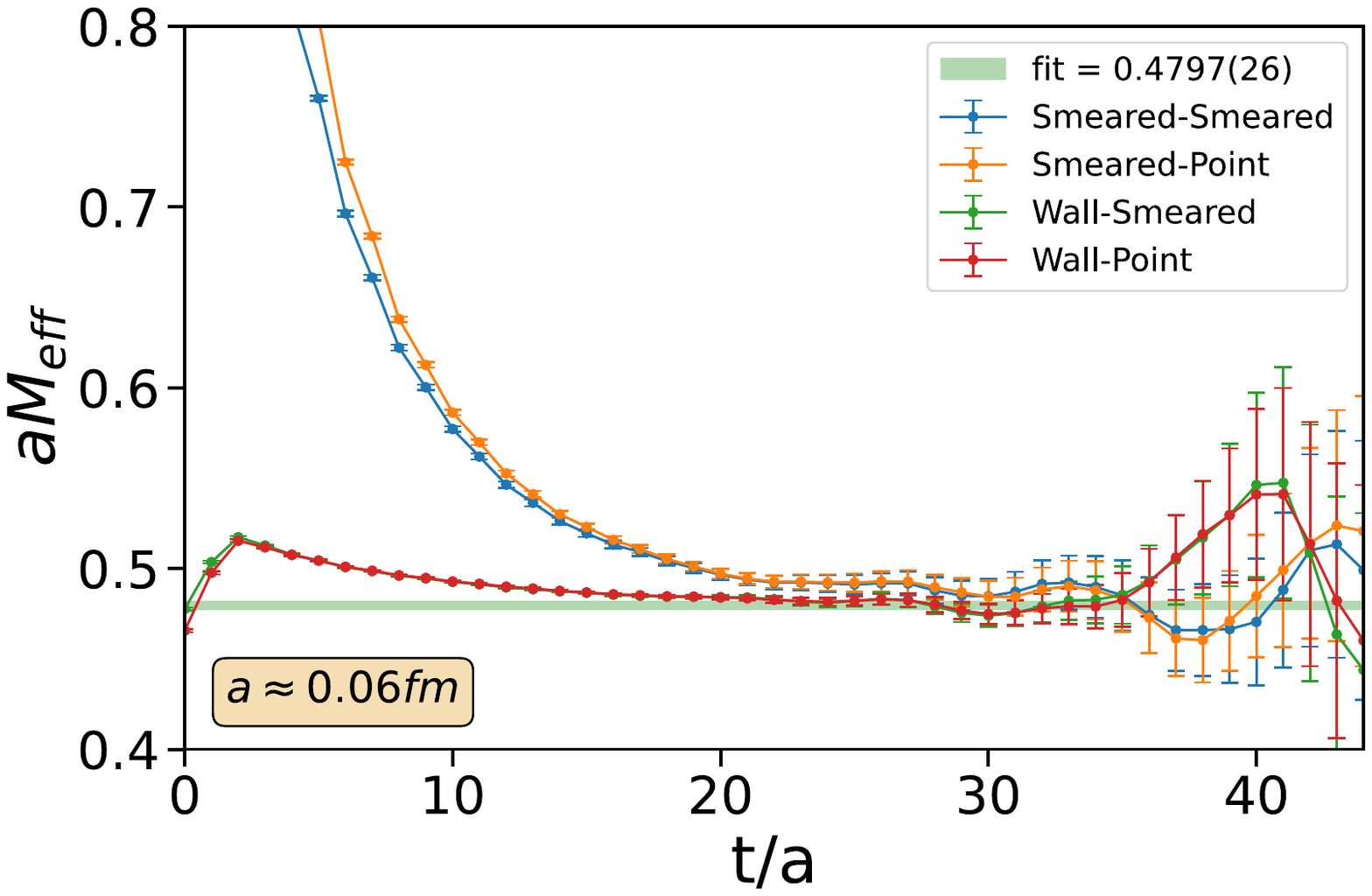}
		\caption{Effective mass plots of the $\Omega$ baryon from two-point correlation functions using Gaussian-smeared and corner-wall sources and Gaussian-smeared and point sinks at two lattice spacings ($a = 0.06,0.09$ fm).  A weighted average over a few time slices has been applied to the correlation functions to suppress the effects of oscillations from negative-parity states~\cite{Lin:2020wko}.\label{fig_omega_a0609}}
	\end{center}
\end{figure}

For each operator, we use four different source-sink constructions: the corner-wall source (see Eq.~(2.1) in Ref.~\cite{Lin:2019pia} for the definition) with the point and Gaussian-smeared sinks and the Gaussian-smeared source with the point and Gaussian-smeared sink. 
Even though correlators constructed from corner-wall sources have the best signal-to-noise ratios, the correlators we construct with corner-wall sources are not positive definite because correlators constructed with corner-wall sinks have unfavorable signal-to-noise scaling in volume and cannot be used in practice. 
So we supplement corner-wall source with Gaussian-smeared sources to constrain the excited state contamination.
In preliminary fits, we have analyzed the correlators constructed from the operator that interpolates to a single $\Omega$-baryon taste for simplicity. We also
apply a weighted average over a few time slices~\cite{Lin:2020wko}
before fitting to alleviate the excited-state contamination from negative-parity states. 
We then perform Bayesian fits to all four correlators on each ensemble with two positive-parity states and one negative-parity state to constrain the left-over negative-parity state contribution. Fit results from the two finest lattice spacings are shown in Fig.~\ref{fig_omega_a0609}.

\begin{SCfigure}
		\includegraphics[width=0.49\textwidth]{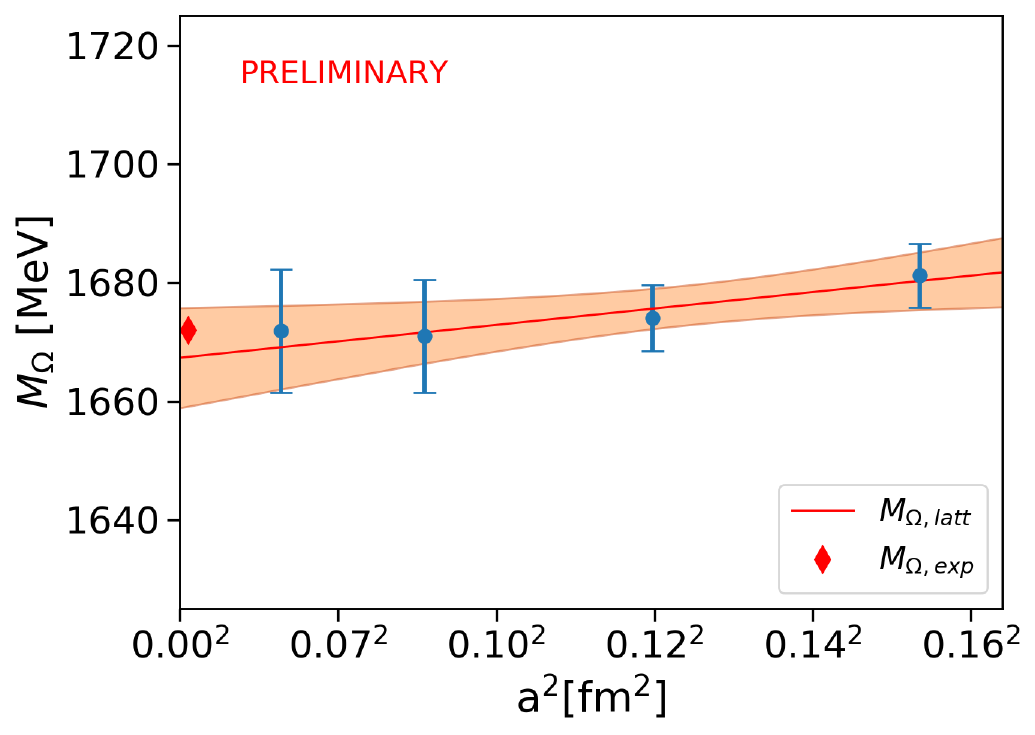}
		\caption{A preliminary continuum extrapolation of the $\Omega$-baryon mass from measurements on ensembles ranging from 0.15 fm down to 0.06 fm using Eq.~\eqref{eq_cont}.
				The lattice spacings are set using $f_\pi$ from Ref.~\cite{Bazavov:2017lyh}.
				The continuum-extrapolated value gives good agreement with the experimental measurement $M_{\Omega, \text{exp}}$.
			\label{fig_omega_cont}}
\end{SCfigure}

Finally, we perform a preliminary continuum extrapolation of the $\Omega$-baryon mass as a consistency check. Here we follow the same procedure used in our earlier work~\cite{Hughes:2019ico} and fit the masses from different lattice spacings, $M_{\text{lat.},\Omega}(a)$, to the functional form
\begin{align}
	M_{\text{lat.},\Omega} = M_{\Omega}\big(1 + c_1(a\Lambda_\text{QCD})^2 + c_2(a\Lambda_{QCD})^4\big)
	\label{eq_cont}
\end{align}
where we choose $\Lambda_\text{QCD}=200$~MeV, $c_1$ is unconstrained, the prior on $c_2$ is $0(1)$, and the lattice spacings are determined using the mass independent scheme with $f_\pi$~\cite{Bazavov:2017lyh}. The result is shown in Fig.~\ref{fig_omega_cont}. The final continuum-extrapolated mass is $M_{\Omega} = 1668(8)$~MeV, which is consistent with the PDG result within a one sigma uncertainty.
We are currently working on increasing statistics on all ensembles to further reduce the uncertainties.

\section{Conclusion}

The effort to determine the $w_0$ scale on all MILC HISQ ensembles is ongoing. The combinations of two flows (Wilson and Symanzik) and three observables (Wilson, Symanzik and clover) have been carried out on all existing MILC (2+1+1)-flavor HISQ ensembles and the final error analysis is in progress. The $\Omega$-baryon mass is being computed on the physical pion mass ensembles for the absolute scale setting. Early analysis indicates that an error of well below 0.5\% may be achievable.

\section*{Acknowledgements}

Computations for this work were carried out in part with resources provided by the USQCD Collaboration, the National Energy Research Scientific Computing Center (NERSC) which are funded by the Office of Science of the U.S.\ Department of Energy
as well as resources provided under the  ASCR Leadership Computing Challenge (ALCC) program at the Argonne Leadership Computing Facility, a U.S. Department of Energy (DOE) Office of Science user facility at Argonne National Laboratory and is based on research supported by the U.S. DOE Office of Science-Advanced Scientific Computing Research Program, under Contract No. DE-AC02-06CH11357.
This work used Ranch tape storage system at Texas Advanced Computing Center (TACC) at The University of Texas at Austin through allocation MCA93S002: Lattice Gauge Theory on Parallel Computers from the Advanced Cyberinfrastructure Coordination Ecosystem: Services \& Support (ACCESS) program~\cite{ACCESS}, which is supported by National Science Foundation grants Nos. 2138259, 2138286, 2138307, 2137603, and 2138296.
Computations on the Big Red II+, Big Red 3 and Big Red 200 supercomputers were supported in part by Lilly Endowment, Inc., through its support for the Indiana University Pervasive Technology Institute.
The parallel file system employed by Big Red II+ was supported by the National Science Foundation under Grant No.~CNS-0521433.
Small-scale computations utilized 
computational resources and services provided by the Institute for Cyber-Enabled Research at Michigan State University.

This work was supported in part by the U.S.~Department of Energy, Office of Science, under Awards
No.~DE-SC0010120 (S.G.),
No.~DE- AC02-07CH11359 (A.V.G and A.S.K), 
No.~DE-SC0011090 (W.I.J. and Y.L.),
No.~DE-SC0021006 (W.I.J.),
No.~DE-SC0015655 (A.X.K.), and
the Funding Opportunity Announcement Scientific Discovery through Advanced Computing: High Energy Physics, LAB 22-2580; 
by the National Science Foundation under 
Grants Nos.~PHY20-13064 and PHY23-10571 (C.E.D.),
No.~PHY18-12332 (A.B.)
and No.~PHY-2019786 (Y.L.) ; 
by the Simons Foundation under their Simons Fellows in Theoretical Physics program (A.X.K.);   
by SRA (Spain) under Grant No.\ PID2019-106087GB-C21 / 10.13039/501100011033 (E.G.);
by the Junta de Andalucía (Spain) under Grants No.\ FQM-101, A-FQM-467-UGR18 (FEDER), and P18-FR-4314 (E.G.). 
This document was prepared by the Fermilab Lattice, and MILC Collaborations using the resources of the Fermi National Accelerator Laboratory (Fermilab), a U.S. Department of Energy, Office of Science, HEP User Facility.
Fermilab is managed by Fermi Research Alliance, LLC (FRA), acting under Contract No.~DE- AC02-07CH11359.

\end{document}